\documentclass[aps,prl,twocolumn,amssymb,superscriptaddress,showpacs]{revtex4-1}

\usepackage{graphicx,amsmath,amssymb,color,epsf,times}

\begin{document}
\title[]{Thermodynamic optimality of glycolytic oscillations}
\author{Pureun Kim}
\affiliation{Korea Institute for Advanced Study, Seoul 02455, Korea}
\author{Changbong Hyeon}
\thanks{hyeoncb@kias.re.kr}
\affiliation{Korea Institute for Advanced Study, Seoul 02455, Korea}


\begin{abstract}
Temporal order in living matters reflects the self-organizing nature of dynamical processes driven out of thermodynamic equilibrium. 
Because of functional reason, the period of a biochemical oscillation must be tuned to a specific value with precision; however, according to the thermodynamic uncertainty relation (TUR), the precision of oscillatory period is constrained by the thermodynamic cost of generating it. 
After reviewing the basics of chemical oscillations using the Brusselator as a model system, we study the glycolytic oscillation generated by octameric phosphofructokinase (PFK), which is known to display a period of several minutes. 
By exploring the phase space of glycolytic oscillations, we find that the glycolytic oscillation under the cellular condition is realized in a cost effective manner. 
Specifically, over the biologically relevant range of parameter values of glycolysis and octameric PFK, the entropy production from the glycolytic oscillation is minimal when the oscillation period is (5 -- 10) minutes. 
Further, the glycolytic oscillation is found at work near the phase boundary of limit cycles, suggesting that a moderate increase of glucose injection rate leads to the loss of oscillatory dynamics, which is reminiscent of the loss of pulsatile insulin release resulting from elevated blood glucose level. 
\end{abstract}


\maketitle

From the perspective of thermodynamics on the macroscale, it is clear that 
living organisms are sustained far from thermodynamic equilibrium in that energy and material currents are supplied to the systems and dissipated to the environment \cite{Bustamante2005PhysicsToday}. 
For human consisting of $\sim 37$ trillions of cells, 
the energy consumption is $\sim 100$ W. 
At the cellular scale, this amounts to as much as $\sim 10^9$ $k_BT$ of energy being consumed per second, 
which drives a vast number of cellular processes.  
Among them, biochemical oscillations, exemplified with circadian rhythm \cite{Rust2011Science,Rust2007Science} and the cell cycle \cite{NovakTyson1993,TysonNovak2001,CellCycle_cell}, are the temporal dissipative structures that emerge from molecular components and their dynamics in nonequilibrium \cite{Prigogine78Science,goldbeter2018PTRSA}.

The full catabolism of glucose under aerobic condition is the primary source of ATP; one glucose molecule could yield as many as 30-32 ATP molecules. 
Besides its utmost importance along the metabolic pathway, 
the sustained oscillation of substrate and product concentrations with time period of $\sim$ (5 -- 10) min observed in the glycolysis of yeast, muscle, and pancreatic $\beta$ cell is a phenomenon of great interest on its own \cite{duysens1957BBA,boiteux1975PNAS,Goldbeter1976Annual,tornheim1997Diabetes,bertram2004BJ}. 
Studies carried out on yeast indicate that among a host of glycolytic enzymes,  
the allostery of phosphofructokinase (PFK), which uses ATP hydrolysis to catalyze the phosphorylation of fructose-6-phosphate (substrate, $S$) into fructose-1,6-biopohosphate (product, $P$), was identified to lie at the core of the endogenous oscillatory mechanism \cite{hess1971ARB,boiteux1975PNAS}. 
The binding of product to the allosteric site induces a cooperative transition of the PFK enzyme from its inactive to active state, introducing a nonlinear positive response to the biochemical circuit.  
The product-activated allosteric regulation of PFK enzyme becomes the source of the glycolytic oscillation.  

In this paper, we study the phosphofructokinase (PFK) model for glycolytic oscillations, and discuss how optimal this energy consuming process is in achieving its desired oscillation period at the expense of thermodynamic cost, which we quantify using the thermodynamic uncertainty relation (TUR),  a recently discovered thermodynamic principle \cite{barato2015PRL,Gingrich2016PRL,horowitz2019NaturePhys,hasegawa2019PRL}.     
In the section \emph{Methods}, 
we briefly review the TUR for dynamical processes displaying temporal oscillations. 
Before studying the PFK model of glycolytic oscillation, we review the Brusselator, a model chemical processes exhibiting oscillations, and analyze it in detail at varying parameter values by means of the phase diagram to identify the optimal condition of the process in light of TUR. 
 The \emph{Results} section covers the allosteric model of octameric PFK, dynamic phase diagram, and the mass balance equations of PFK-catalyzed substrate and product giving rise to the glycolytic oscillations. We assess the optimality of glycolytic oscillations realized by octameric PFK in terms of the period and fluctuations of the oscillations and the associated thermodynamic cost of generating such dynamics, which can be quantified using TUR.  
 Finally, we will underscore the optimality of glycolytic oscillations emerging from octameric PFK by comparing it with the dynamics anticipated for hypothetical non-octameric PFK model.        

\section{Methods} 
\subsection{Thermodynamic uncertainty relation for oscillatory processes}
Biological processes are inherently stochastic due to fluctuations in the cellular environment. 
Suppose that time trajectories of a dynamical process generated from a constant thermodynamic drive are available. 
Then, in order to gain knowledge about the dynamics from a given trajectory with higher precision, it is required to analyze a longer trajectory; however, generation of a longer time trajectory incurs larger thermodynamic cost. 
According to the thermodynamic uncertainty relation (TUR), there are trade-offs between the total entropy production 
$\Delta S_{\text{tot}}(t)$ from the process generated under a constant drive and the precision of a time integrated \emph{current-like} observable $X(t)$ to probe the dynamical process generated in nonequilibrium.  
Furthermore, the product between $\Delta S_{\text{tot}}(t)$ and the squared relative uncertainty, $\text{Var}[X(t)]/\langle X(t) \rangle^2$, defined as the  \emph{uncertainty product} $\mathcal{Q}$, is bounded below by twice the Boltzmann constant ($2k_{\text{B}}$):
\begin{equation}
\mathcal{Q} \equiv \Delta S_{\text{\text{tot}}}(t)\frac{\text{Var}[X(t)]}{\langle X(t) \rangle ^2} \geq 2k_{\text{B}}. 
\label{eqn:TUR}
\end{equation}
Because of the lower bound, there is a minimal thermodynamic (entropic) cost to generate a process with a certain precision, or that the precision of the process is bounded by the thermodynamic cost being expended.  
The relation applies to current-like output observables, with odd parity under time reversal satisfying $X(t)=-X(-t)$ \cite{hasegawa2019PRL}, that arise from dynamic processes that can be modeled using continuous time Markovian dynamics 
in discrete network or in continuous space under \emph{time-independent} driving. 
There have been further generalizations of the relation \cite{Proesmans:2017,horowitz2019NaturePhys,hasegawa2019PRL,lee2019PRE,agarwalla2018PRB,lee2018PRE,potts2019PRE,koyuk2018JPA} including the those under time-dependent driving \cite{koyuk2018JPA,koyuk2019PRL,koyuk2020PRL}. 
Here, we confine ourselves to the original relation (Eq.\ref{eqn:TUR}) since the oscillatory dynamics considered in this study are generated under time-independent driving. 
Throughout the paper, we set the Boltzmann constant $k_B$ to unity for simplicity of the expressions. 

There have been several applications of TUR to characterize biological processes \cite{pietzonka2016JSM,Hwang2018JPCL,uhl2018PRE,Song2020JPCL,Pineros2020PRE,marsland2019JRSI,Song2021JCP}. For biological motors that move along filaments, the net displacement of a motor $\Delta x(t)=x(t)-x(0)=-\Delta x(-t)=x(0)-x(-t)$ satisfies the odd parity, with time translation symmetry, and may be selected as a proper output observable to probe the functional dynamics of the motor. 
For dynamical processes displaying temporal oscillations, 
the time duration of the process, $t$, can be employed as a current-like observable to monitor the progress of {\color{blue}$n$} biochemical oscillations, i.e., $X(t)=t$.  
In fact, it has recently been shown through the large-deviation theory that in the limit of long times and large currents TUR holds for the first passage times \cite{gingrich2017PRL} as well, so that a dynamical process consisting of $n(\gg 1)$ oscillations that has occurred for a sufficiently long time duration $t$ satisfies the following relation: 
\begin{align}
\mathcal{Q}=\Delta S_{\rm tot}(t)\frac{\langle\delta t^2\rangle}{\langle t\rangle^2}\geq 2k_B. 
\end{align}
The time duration, $t$, is decomposed as $t=\sum_{i=1}^n T_i$ where $n$ is the number of oscillations for time $t$ and $T_i$ is a stochastic variable representing the period of $i$-th oscillation. 
Then, the first and second moments of the time duration are written as 
\begin{align}
\langle t\rangle =\sum_{i=1}^n\langle T_i\rangle = n \langle T\rangle
\label{eqn:mean}
\end{align}
and 
\begin{align}
\langle t^2\rangle &=\sum_{i=1}^n\langle T_i^2\rangle +2\sum_{i>j}\langle T_iT_j\rangle\nonumber\\
&=n\langle T^2\rangle +n(n-1)\langle T_iT_j\rangle\nonumber\\
&=n\langle T^2\rangle -n\langle T\rangle^2+\langle t\rangle^2
\label{eqn:second}
\end{align}
where $\langle T\rangle\equiv (1/n)\sum_{i=1}^nT_i$ is the mean oscillatory period. 
For renewal processes, the independence of two oscillatory periods, $\langle T_iT_j\rangle=\langle T_i\rangle \langle T_j\rangle=\langle T\rangle^2$, can be used, giving rise to the last line of Eq.\ref{eqn:second}. \cite{marsland2019JRSI}
From Eqs.~\ref{eqn:mean} and \ref{eqn:second}, 
one obtains $\langle t^2\rangle-\langle t\rangle^2=n\langle T^2\rangle-n\langle T\rangle^2=n\times{\rm Var}(T)$, which indeed holds for the Brusselator studied here (see Fig.~S1).  
Even in the presence of a finite correlation between oscillatory dynamics, 
$\langle \delta T_i \delta T_j\rangle\simeq 0$ holds as long as the time gap corresponding to $|i-j|$ is greater than the correlation time. 
Alternatively, in 1990s Schnitzer and Block \cite{schnitzer1995statistical} showed for processive enzymatic processes that the number of net catalytic events that have occurred for time $t$, $n(t)$, is related to the enzymatic cycle time $\tau$, which corresponds to the oscillatory period $T$ in this work, namely $\tau=T$, as 
$\lim_{t\rightarrow\infty}\langle \delta n(t)^2\rangle/\langle n(t)\rangle=\langle \delta \tau^2\rangle/\langle \tau\rangle^2=\langle\delta T^2\rangle/\langle T\rangle^2$. 

Taken together, the uncertainty product $\mathcal{Q}$ for processes demonstrating temporal oscillations as follows \cite{marsland2019JRSI,cao2015NatPhys,FrankPRL2007}: 
\begin{align}
\mathcal{Q}&=\Delta S_{\text{tot}}(t)\frac{\langle  t^2\rangle-\langle  t\rangle^2}{\langle  t\rangle^2}\nonumber\\
&\simeq \left[\frac{\Delta S_{\text{tot}}(t)}{n\langle T\rangle}\right]\frac{\text{Var}(T)}{\langle T\rangle}\nonumber\\
&=\dot{S}_{\text{tot}}\frac{\text{Var}(T)}{\langle T\rangle}\nonumber\\
&=\Delta S_\text{cyc}\frac{\text{Var}(T)}{\langle T\rangle^2}\geq 2k_B, 
\label{eqn:TUR_T}
\end{align}
where $\text{Var}(T)=\langle(\delta T)^2\rangle=\langle T^2\rangle-\langle T\rangle^2$. 
$\dot{S}_{\text{tot}}=\Delta S_{\text{tot}}(t)/n\langle T\rangle=\Delta S_{\text{tot}}(t)/t$ is the rate of entropy production, and the entropy production per cycle can be defined as $\Delta S_{\text{cyc}}\equiv \Delta S_{\text{tot}}(t)/n=\dot{S}_{\text{tot}}\times \langle T\rangle$. 
The inequality in the last line of Eq.\ref{eqn:TUR_T} is identical to the expression $\mathcal{N}\equiv \langle T\rangle^2/D\leq \Delta S_{\text{cyc}}/2$ ($k_B=1$) of Marsland \emph{et al.}'s \cite{marsland2019JRSI,marsland2017RPP} 
where $D\equiv\text{Var}(T)=\langle (\delta T)^2\rangle$ and $\mathcal{N}$ corresponds to the number \emph{coherent} oscillations.  
If the uncertainty product $\mathcal{Q}$ evaluated for a process displaying  temporal oscillations is close to the theoretical minimum $2k_B$, one could presume that the process is close to its optimal condition where the thermodynamic cost of generating an oscillatory dynamics with a certain temporal precision is minimal. 

For a given oscillatory process arising from a set of kinetic equations, 
it is straightforward to evaluate the mean ($\langle T\rangle$) and variance of oscillation period ($\text{Var}(T)$) in Eq.\ref{eqn:TUR_T} from the time trajectories of dynamical variables, ${\bf x}(t)$, that satisfy ${\bf x}(t)\simeq {\bf x}(t+T)$.  
To obtain the total entropy production rate from ${\bf x}(t)$, one can utilize the evolution equation of probability density, $P({\bf x},t)$, namely the Fokker-Planck (FP) equation. 
The FP equation is obtained from a corresponding set of chemical master equations (CME) at a finite volume ($\Omega$) (see SI): 
\begin{align}
{\frac{\partial P({\bf x},t)}{\partial t}} &= -\vec{\nabla}\cdot\left[{\bf H}\cdot\vec{F}({\bf x})-{\bf D}({\bf x})\cdot\vec{\nabla}\right]P({\bf x},t)\nonumber\\
&= -\vec{\nabla}\cdot\vec{J}({\bf x},t),
\label{eqn:FP}
\end{align}
where ${\bf D}({\bf x})(=k_BT {\bf H}({\bf x}))$ is the $\Omega$-dependent diffusion tensor with ${\bf H}({\bf x})$ being the motility tensor, and 
\begin{align}
\vec{J}({\bf x},t) = {\bf H}({\bf x})\cdot\vec{F}({\bf x})P({\bf x},t)-{\bf D}({\bf x})\cdot\vec{\nabla} P({\bf x},t)
\label{eqn:current}
\end{align}
is the probability current, $\vec{F}({\bf x},t)$ is the driving force vector. 
The total entropy production rates, contributed by both the system and reservoir, are obtained by 
averaging the corresponding \emph{trajectory-based} entropy production rates over the probability density $P({\bf x},t)$, 
\begin{align}
\dot{S}_\text{tot}(t)=\langle \dot{s}_\text{tot}(t)\rangle=\int \dot{s}_{\text{tot}}(t)P({\bf x}(t),t)d{\bf x},  
\label{Stot}
\end{align}
where $\dot{s}_\text{tot}(t)=\dot{s}_\text{sys}(t)+\dot{s}_\text{res}(t)$ with $s_{\text{sys}}(t)=-\log{P({\bf x}(t),t)}$ and $\dot{s}_{\text{res}}(t)(=\vec{F}({\bf x},t)\cdot \dot{\bf x})/T_r$ where $T_r$ is the temperature of the reservoir \cite{seifert2005PRL,qian2001PRE,QianEntropyPRE2010,OliveiraPRE}. 
Together with the expression of probability current (Eq.\ref{eqn:current}), Eq.\ref{Stot} yields \cite{seifert2005PRL} 
\begin{equation}
\dot{S}_{\text{\text{tot}}}(t)=\int \frac{\vec{J}^{\intercal}({\bf x},t)\cdot{\bf D}^{-1}({\bf x})\cdot\vec{J}({\bf x},t)}{P({\bf x},t)}d{\bf x}.  
\label{eqn:S}
\end{equation}
The entropy production rate at steady state, obtained from $J^{\text{ss}}({\bf x})$ and $P^{\text{ss}}({\bf x})$, allows us to evaluate the uncertainty product $\mathcal{Q}$ at steady state (Eq.\ref{eqn:TUR_T}). 

Cautionary remarks are in place regarding the use of Eq.\ref{eqn:S} to evaluate the entropy production, which is derived from the Fokker-Planck equation describing the time evolution of the probability density for the dynamic variables ${\bf x}$. 
In this study, we employed a widely adopted strategy of approximating the chemical master equations (CME) for the Brusselator and the PFK model for glycolytic oscillation to the corresponding Fokker-Planck equations via the van Kampen's linear noise approximation (LNA or $\Omega$-expansion) (see SI) \cite{vanKampen,schuster2016stochasticity,Qian_PNAS,wang2008PNAS,xiao2008JCP,cao2015NatPhys}. 
However, consistency between CME and LNA approach has recently been questioned in the context of stochastic thermodynamics \cite{grima2010JCP,grima2011JCP,horowitz2015JCP}. 
If the system size parameter $\Omega$ is too small, not only the accuracy of the approximation becomes questionable \cite{grima2010JCP,grima2011JCP}, but the entropy production rate calculated by employing Eq.\ref{eqn:S} is also bound to \emph{underestimate} the true value \cite{horowitz2015JCP}. 
Since the system size we have chosen for the simulation ($\Omega=1600$) is large enough that the approximation is essentially taken in the regime where the discrepancy between the entropy productions calculated from CME and from Fokker-Planck approach should not be significant.
Another possible cause of underestimation of entropy production arises when one adopts the coarse-graining \cite{yu2021PRL} or the projection of dynamics to slow degrees of freedom \cite{ZwanzigBook,vandenBroeck2010PRE}. 
For the cases of the Brusselator and the glycolytic oscillation studied here, their reaction dynamics are defined with two stochastic variables ($x$ and $y$ for the Brusselator; $[S]$ and $[P]$ for the glycolytic oscillation), we probe both stochastic variables that are slowly varying with time and faithfully represent the oscillatory dynamics.

\begin{figure*}[t]
\includegraphics[width=1.0\textwidth]{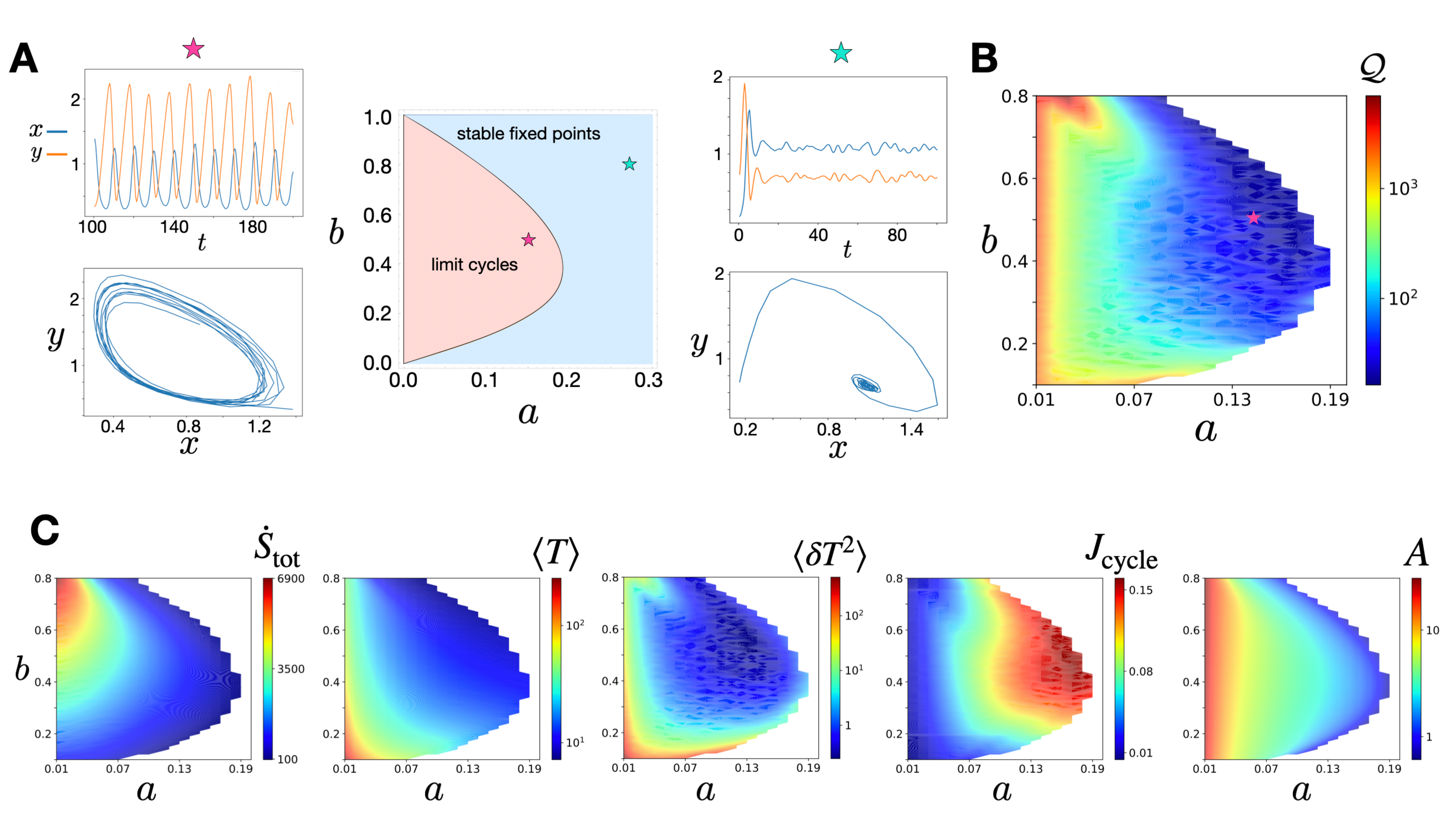}
\caption{{\bf A}. Dynamic phase diagram of the Brusselator. 
The steady state oscillation of limit cycle occurs in the pink region, left side of the phase boundary.  
The blue region marked with ``stable fixed points" displays no oscillations at steady states. 
Two exemplary trajectories, one (red star) from the region of limit cycles and the other (cyan star) from the region of stable fixed points are shown. 
The corresponding phase planes with the vector fields are depicted in Fig.~S2. 
{\bf B}. The 2D diagram of $\mathcal{Q}(a,b)$ corresponding to the phase region of limit cycles.  
{\bf C}. The 2D diagrams for other quantities, $\dot{S}_{\text{tot}}$, $\langle T\rangle$, $\langle(\delta T)^2\rangle$, $J_\text{cycle}$, and $A$, as a function of $a$ and $b$. 
}
\label{fig:TURofbrussel}
\end{figure*}

\subsection{Brusselator}
The Brusselator, a model for autocatalytic reactions that display sustained oscillatory dynamics in certain range of parameters, offers all the ingredients of oscillatory dynamics required to learn the dissipation and precision of glycolytic oscillations to be studied. 

The model forms an open system, consisting of two chemical compounds $X$ and $Y$ reacting each other and being depleted while other source compounds $A$ and $B$ are constantly supplied to the system at fixed concentrations.  
The reaction scheme for Brusselator is
\begin{align} 
A&\xrightarrow{k_1} X, \nonumber\\
B&\xrightarrow{k_2} Y, \nonumber\\
2X+Y&\xrightarrow{k_3} 3X, \nonumber\\
X&\xrightarrow{k_4} \phi. 
\end{align}
Properly non-dimensionalized (see SI),
the rate equations at the limit of an infinite volume $(\Omega \rightarrow  \infty)$ is given as 
\begin{align}
\frac{dx}{dt} &= f(x,y)=a - x+x^2y,\nonumber\\
\frac{dy}{dt}&=g(x,y)=b-x^2y,
\label{eqn:Brusselator}
\end{align}
where $x$, $y$, $a$, and $b$ are the non-dimensionalized concentrations of chemical species $X$, $Y$, $A$ and $B$, respectively.  
The equations for the concentrations of $x$ and $y$ are nonlinear; thus when $x$ and $y$ are expanded around a fixed point $(x^{\ast},y^{\ast})$ that satisfies $f(x^{\ast},y^{\ast})=0$ and $g(x^{\ast},y^{\ast})=0$ such that $x=x^\ast+\delta x$ and $y=y^\ast+\delta y$, it yields a set of linearized equations  
\begin{align}
\delta\dot{{\bf x}}=\mathcal{J}(x^\ast,y^\ast)\cdot\delta{\bf x}. 
\end{align}
where $\delta{\bf x}\equiv(\delta x,\delta y)^{\intercal}$ and 
\begin{align}
\mathcal{J}(x^{\ast},y^{\ast})&=\begin{bmatrix} 
f_x & f_y \\
g_x & g_y \nonumber 
\end{bmatrix}
\end{align}
is the Jacobian matrix evaluated at $(x^\ast,y^\ast)=(a+b,b/(a+b)^2)$, 
where $f_x\equiv \partial_xf(x,y)|_{(x,y)=(x^\ast,y^\ast)}$. 
Since the fluctuation $\delta{\bf x}$ is expected to change with time as $\delta{\bf x}\sim e^{\lambda t}$, the stability of the fixed point is determined by the real parts of eigenvalues ($\lambda$) obtained from the characteristic equation $\det{\left[\lambda\mathcal{I}-\mathcal{J}(x^{\ast},y^{\ast})\right]}=0$ where $\mathcal{I}$ is the identity matrix. 
The eigenvalues are:  
\begin{align}
\lambda=\frac{1}{2}(\tau\pm\sqrt{\tau^2-4\Delta})
\end{align}
with $\tau=-(a+b)^2-(a-b)/(a+b)$ and $\Delta=(a+b)^2$. 
Since $\Delta>0$ for any $a$ and $b$, the sign of ${\bf Re}(\lambda)$ is determined entirely by the sign of $\tau$. 
The condition of $\tau=0$, leading to  $b^3+3ab^2+(3a^2-1)b+a^3+a=0$, determines the phase boundary (Fig.\ref{fig:TURofbrussel}A). 
The set of parameters $(a,b)$ belonging to the blue region of phase diagram (Fig.\ref{fig:TURofbrussel}A) leads to $\tau<0$, then the fixed points are stable, and the time trajectory of $(x(t),y(t))$ converges to $(x^\ast,y^\ast)$ (see the lower rightmost panel of Fig.\ref{fig:TURofbrussel}A).  
On the other hand, the region colored in red ($\tau>0$) yields 
unstable fixed points that produce limit cycles (see Fig.~S2A and B for the vector fields $(\dot{x}(t),\dot{y}(t))$ depicted for $(a,b)$ leading to unstable and stable fixed points).  
For the case of 2D phase plane, existence of limit cycles is always guaranteed by the Poincar{\'e}-Bendixson theorem \cite{strogatz2018nonlinear}.

\begin{figure*}[t]
\includegraphics[width=0.9\linewidth]{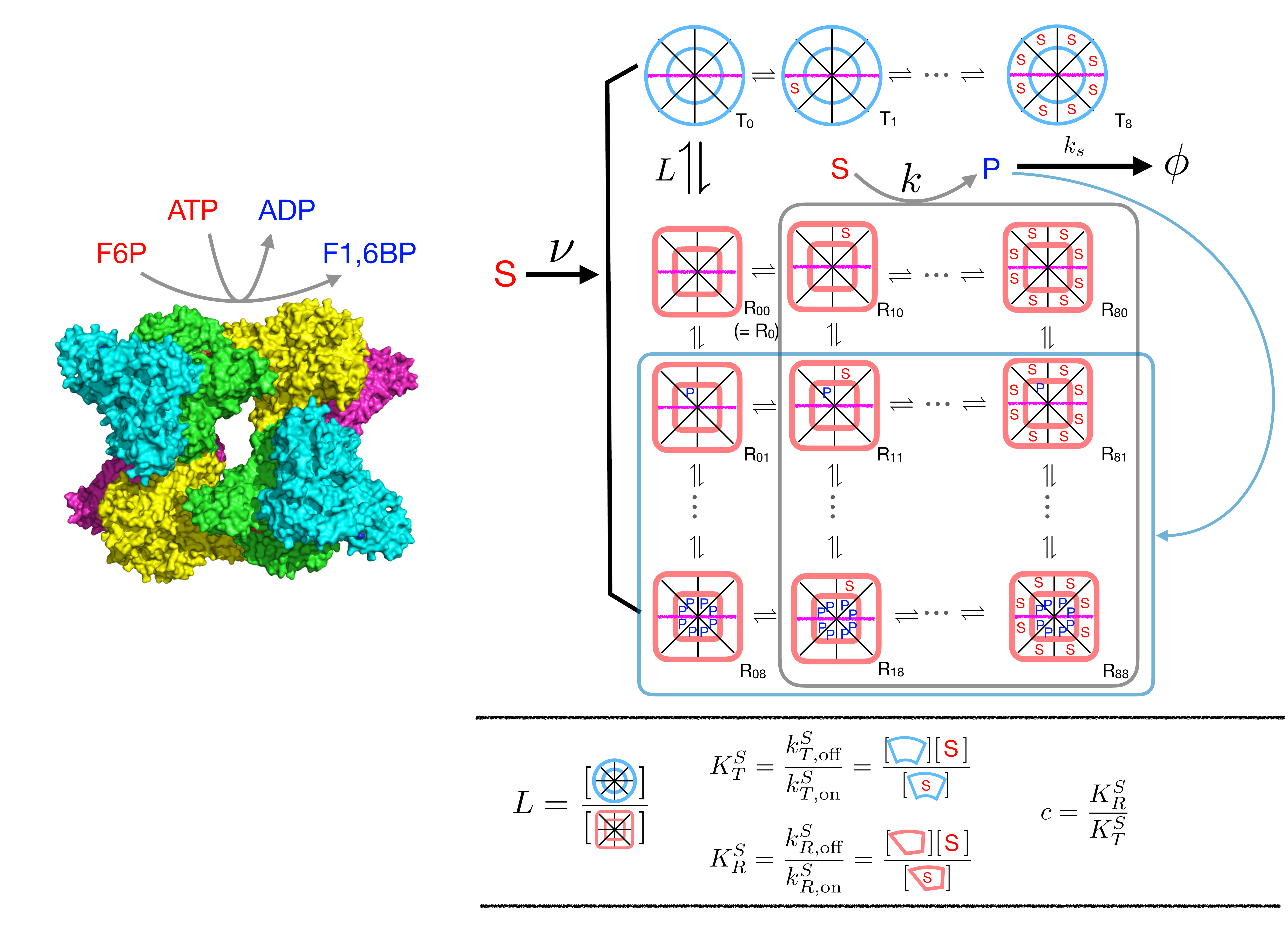}
\caption{Allosteric model for PFK in glycolysis. 
The PFK1 is an oligomeric enzyme, which takes different oligomeric state depending on the organism. 
Depicted is the octameric structure (dimer of tetramer) of PFK1 in $R$ state (PDB code 4U1R \cite{kloos2015BJ}). 
Each subunit has substrate ($S=$ F6P) 
and product ($P=$ FBP) binding site, the latter of which allosterically regulate the enzymatic activity of the enzyme. 
The PFK model of glycolytic oscillation assumes an open thermodynamic system where glucose (or substrate) molecules are injected at the rate $\nu$ and the product ($P$) is drained at $k_s$. Depicted is the octameric form of enzyme with $n=8$ binding sites for substrate and product. 
At low substrate concentration, the enzyme is in the \emph{tense} ($T$) state. But, with increasing substrate concentration, transition to the $R$ state occurs, increasing the catalytic activity of enzyme. 
}
\label{fig:allostericP}
\end{figure*}

For a system with a finite volume, $\Omega$, the FP equation derived from CME \cite{Qian_PNAS} for the Brusselator is obtained with the driving force vector $\vec{F}(x,y)$ and diffusion tensor ${\bf D}(x,y)$. 
\begin{align}
\vec{F}(x,y)&=\begin{bmatrix} 
a-x+x^2y \\
b-x^2y 
\end{bmatrix}
+\frac{1}{2\Omega}\begin{bmatrix}
-1/2-2xy+x^2/2 &\\
2xy-x^2/2
\end{bmatrix},\nonumber\\
{\bf D}(x,y)&=\frac{1}{2\Omega}\begin{bmatrix}
a+x+x^2y & -x^2y \\
-x^2y & b+x^2y 
\end{bmatrix}.
\label{eqn:FD}
\end{align}
The $(a,b)$-dependent steady state probability distribution $P_\text{ss}(x,y)$, obtained from an ensemble of trajectories generated using the simulations based on Gillespie's algorithm, allows 
us to calculate the $\vec{J}_{\text{ss}}$ (Eq.\ref{eqn:FP}), and hence $\dot{S}_{\text{\text{tot}}}$ at the steady state using Eq.\ref{eqn:S}.

In the parameter range of $(a,b)$ yielding unstable fixed points ($\tau>0$), 
we have computed $(a,b)$-dependent 2D diagrams of various quantities at $\Omega=1600$:  
$\dot{S}_\text{tot}$, $\langle T\rangle$, $\langle\delta T^2\rangle$, amplitude of oscillations ($A$), and the integral current $J_{\text{cycle}}=\oint \vec{J}_{\text{ss}}\cdot d\vec{l}\Big/\oint dl$ (Fig~\ref{fig:TURofbrussel}). 
The entropy production rate $\dot{S}_\text{tot}$ show overall positive correlation with 
$\langle T\rangle$, $\langle\delta T^2\rangle$, and $A$, but not with $J_{\text{cycle}}$. 
The larger $J_{\text{cycle}}$ signifies faster oscillations. 
It is noteworthy that the correlation or anti-correlation between the quantities calculated here is relatively clear for the Brusselator, but same is not necessarily true for glycolytic oscillator (compare Fig.~S3A and B). 
The $\langle\delta T^2\rangle$, displaying a non-monotonic variation, is minimized at a basin of parameter space around $(a,b)\approx (0.13,0.55)$. 
The product of $\dot{S}_{\text{tot}}$ and $\langle \delta T^2\rangle/\langle T\rangle$  gives rise to the 2D diagram of $\mathcal{Q}(a,b)$ (Fig.\ref{fig:TURofbrussel}), indicating that 
$\mathcal{Q}$ is minimized in the vicinity of the phase boundary $(a,b) = (0.14,0.5)$ to $\mathcal{Q} \approx 17$.

Importantly, along a noisy limit cycle over the range of $\Omega$ being varied (see Fig.~S4), $\mathcal{Q}$ is independent of the system size $\Omega$. 
The entropy production is an extensive quantity that linearly increase with $\Omega$ for a given time interval $t$. 
Thus, the entropy production rate scales with the volume as $\dot{S}_\text{tot}\sim \Omega$.\cite{xiao2008JCP,xiao2009JPCB}
Next, the fluctuation of the oscillatory period $\langle (\delta T)^2\rangle$ is proportional to the magnitude of the $\Omega$-dependent diffusion tensor ${\bf D}$, such that $\langle (\delta T)^2\rangle \sim {\bf D}\sim \Omega^{-1}$ as defined in Eq.\ref{eqn:FD}, whereas $\langle T\rangle$ is decided independently from $\Omega$. 
Taken together, the uncertainty product $\mathcal{Q}$ is a quantity independent of $\Omega$, which can also be confirmed using a host of simulations carried out at fixed parameter values with varying $\Omega$ (see Fig.~S4).

 \section{Results}


The allosteric regulation of PFK1 and its substrate and product concentration-dependent enzymatic activity can be formulated using the strategy of Mono-Wyman-Changeaux model \cite{Mono65JMB,Thirumalai19ChemRev}. 
The enzyme PFK1, a oligomeric complex consisting of $n$ catalytic and $n$ regulatory sites to which the substrate and product, respectively, can bind, is equilibrated between tense (inactive, $T$) and relaxed (active, $R$) states, which differ in terms of their conformations and binding affinities to substrate and product (see Fig.\ref{fig:allostericP}). 
$T$ state can only accommodate substrate molecules, and the subscript $i(=0,1,\ldots n)$ in $T_i$ denotes the number of substrates bound to the binding sites. 
On the other hand, $R$ state can accommodate both substrate and product molecules; the two subscripts $i(=0,1,\ldots,8)$ and $j(=0,1,\ldots,n)$ of $R_{ij}$ denote the number of substrates and products bound to the catalytic and regulatory sites, respectively. 
In the absence of substrate, two apo states of PFK1, $R_0(\equiv R_{00})$ and $T_0$ states, are chemically equilibrated with the ratio,  $L=[T_0]/[R_0]$ called allosteric constant, which determines the degree of cooperativity of the enzyme.  
The glucose converting into substate $S$ (fructose-6-phosphate, F6P) via multiple steps is injected at a constant rate $\nu$ while the product $P$ (fructose-1,6-biopohosphate, FBP) either binds exclusively to the allosteric sites of the $R$ state acting as a positive activator, or degrades at a rate of $k_s$. 
The increase of F6P as a result of the catalytic processes of glycolysis in turn increases the amount of FBP, which positively regulates the catalytic activity of PFK1 by promoting the $T$-to-$R$ transitions. 
The nonlinear response of FBP-binding induced autocatalytic activation of PFK1 
generates the glycolytic oscillation \cite{tornheim1988JBC,yaney1995Diabetes}. 

We assume that the binding and unbinding of substrate ($S$) and product ($P$) to and from each binding site of $R$ state occur with the rates $k_{R,\text{on}}^S$, $k_{R,\text{off}}^S$, and $k_{R,\text{on}}^P$, $k_{R,\text{off}}^P$, 
and that only the $S$ can bind/unbind to $T$ state with $k_{T,\text{on}}^S$, $k_{T,\text{off}}^S$. 
Then, the concentration of each state of PFK1 is obtained as follows by assuming a quasi-steady-state approximation \cite{goldbeter1972BJ}. 
\begin{align}
[T_i]&=\binom{n}{i}\left(\frac{[S]}{K_T^S}\right)^i[T_0]
=L\binom{n}{i}\left(\frac{c[S]}{K_R^S}\right)^i[R_0],\nonumber\\
[R_{ij}]&=\binom{n}{i}\left[\alpha([S])\right]^i\binom{n}{j}\left[\gamma([P])\right]^j[R_0]
\label{eqn:conc}
\end{align}
where 
\begin{align}
\alpha([S])&=\frac{[S]}{(K_R^S+k/k_{R,\text{on}}^S)},\nonumber\\
\gamma([P])&=\frac{[P]}{K_R^P},\nonumber
\label{eqn:alpha_gamma} 
\end{align}
$K_{R}^S(=k_{R,\text{off}}^S/k_{R,\text{on}}^S)$ and  $K_{T}^S(=k_{T,\text{off}}^S/k_{T,\text{on}}^S)$ are the binding affinities (dissociation constants) of the substrate to the catalytic site in the $R$ and $T$ states, respectively, whereas $K_R^P$ is the binding affinity of the product to the regulatory site in the $R$ state. 
The parameter $c=K_R^S/K_T^S$ denotes the ratio of the dissociation constants of the substrate from the catalytic sites in $R$ and $T$ states. 
Then, from the concentration of each enzyme state $[T_i]$ and $[R_{ij}]$ (Eq.\ref{eqn:conc}), 
it is straightforward to calculate a binding polynomial $\bar Y$ ($0\leq \bar{Y}\leq 1$), namely the fraction of catalytic sites in the $R$ state bound by the substrate, 
\begin{align}
\bar Y &=\frac{\text{total substrates bound to $R$ state}}{n\times\text{total enzymes}}\nonumber\\
&=\frac{\sum_{i=0}^n\sum_{j=0}^ni[R_{ij}]}{nZ([S],[P])}\nonumber\\
&=\frac{\alpha([S])q([S],[P])}{nZ([S],[P])}
\end{align}
where 
\begin{align}
q([S],[P]) &\equiv n(1+\alpha([S]))^{(n-1)}(1+\gamma([P]))^n[R_0],\nonumber
\label{eqn:q} 
\end{align}
and the total concentration of enzyme $Z([S],[P])$ 
\begin{align}
Z&([S],[P])=\sum_{i=0}^n[T_i]+\sum_{i=0}^n\sum_{j=0}^n[R_{ij}]\nonumber\\
&=\left[L\left(1+\frac{c[S]}{K_R^S}\right)^n+(1+\alpha([S]))^n(1+\gamma([P]))^n\right][R_0]
\end{align}

\begin{figure*}
\includegraphics[width=0.9\linewidth]{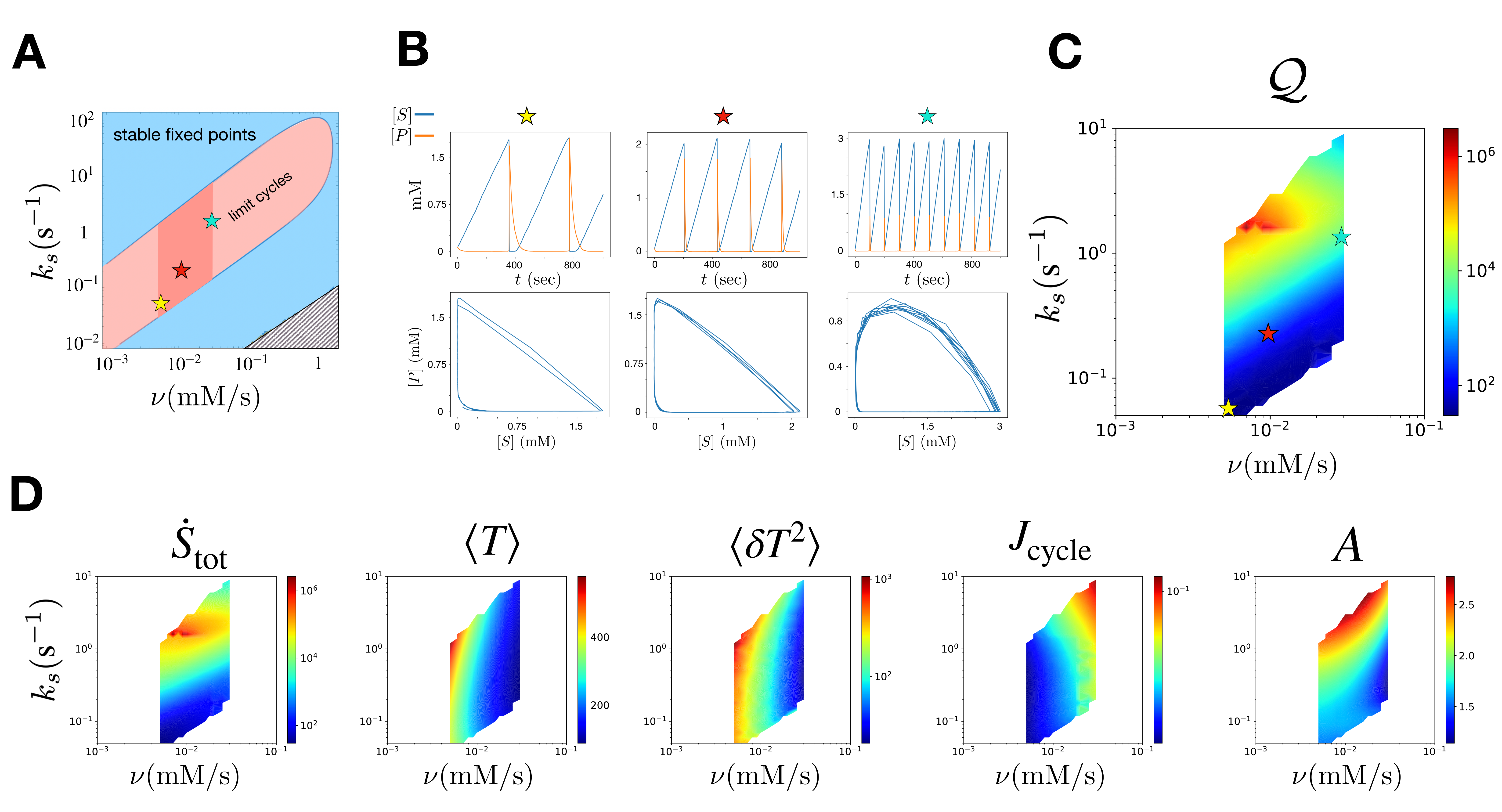}
\caption{
{\bf A}. Phase diagram of glycolytic oscillations as a function of $\nu$ and $k_s$ with 
the allosteric constant ($L=4\times10^9$), catalytic rate ($k=500$ $s^{-1}$) and $c=0.01$.  
No catalysis occurs in the hashed region.  
The simulations were performed with $\nu =0.005-0.03$ mM/s, demarcated with the dark pink region, corresponding to the glucose uptake rate of yeast.
{\bf B}. Three representative trajectories showing oscillatory dynamics of substrate and product concentrations. 
The trajectory marked with yellow star was generated at the parameter values relevant for octameric PFK of yeast giving rise to the oscillatory period of $\sim$ 400 s    \cite{BLANGY1968,HESS1969,goldbeter1972BJ,McCoy2005,cardon1978EJB,Moyer1998} ($L=4\times 10^9$, $k=500$ s$^{-1}$, $\nu=0.005$ mM/s, $k_s=0.05$ s$^{-1}$, $c=0.01$). 
{\bf C}. 2D diagram of uncertainty product, $\mathcal{Q}$ as a function of $\nu$ and $k_s$. 
{\bf D}. $\dot{S}_\text{tot}$, 
$\langle T\rangle$, $\langle(\delta T)^2\rangle$, 
$J_\text{cycle}$, and $A$ calculated as a function of $\nu$ and $k_s$. 
}
\label{fig:TURofglycolysis}
\end{figure*}

Substrates supplied with constant rate $\nu$ bind to $T$ or $R$ state, and is catalyzed by the $R$ state of PFK1 ($R_{ij}$) with a rate $k$ to generate product molecules, 
whereas the products either bind to the regulatory site of $R$ state or depleted from the system at a rate $k_s$.   
The mass action laws of the substrate and product yield a set of coupled nonlinear equations 
\cite{goldbeter1972BJ}: 
\begin{align}
\frac {d[S]}{dt} &=\nu-k\sum_{i=0}^n\sum_{j=0}^ni[R_{ij}]= \nu - k\alpha([S])q([S],[P]),\nonumber\\
\frac {d[P]}{dt} &=k\sum_{i=0}^n\sum_{j=0}^ni[R_{ij}]-k_s[P]= k\alpha([S])q([S],[P]) - k_s [P]. 
\label{eqn:glycolysis}
\end{align}
Here, by specifically considering the yeast PFK which adopts an octameric form ($n=8$), we use the allosteric constant $L=4 \times 10^9$ which is 3 orders of magnitude greater than the value known for the tetramer ($n=4$) \cite{BLANGY1968}. 
Other parameters related with substrate/product binding to each protomer are expected to be identical between tetramer and octamer, thus we take $k_{R,\text{on}}^S = 2000 $/mM$\cdot$s, $K^S_R = 0.05$ mM and $K^P_R = 0.025$ mM \cite{McCoy2005,Moyer1998,BLANGY1968,cardon1978EJB}. 
The ratio of the substrate binding affinities to the $R$ and $T$ states is  
$c=K_R/K_T=0.01$, so that 
the substrate binds preferentially to the $R$ state. 
The ATP hydrolysis time due to ATPase activity is typically $\gtrsim\mathcal{O}(1)$ msec \cite{Gilbert94Biochem}, and hence we set $k=500$ s$^{-1}$. 

Following the same procedure used in the analysis of Brusselator, 
we analyze Eq.\ref{eqn:glycolysis} to calculate phase diagrams as a function of $\nu$ and $k_s$ (Fig.\ref{fig:TURofglycolysis}A) and generate dynamical trajectories of $[S](t)$ and $[P](t)$ (Fig.\ref{fig:TURofglycolysis}B).  
For the values of $(\nu,k_s)$ pertaining to the limit cycles, 
the substrate concentration display saw-tooth like oscillatory pattern in time, and and the product concentration spikes when the substrate concentration falls, which generates a loop in the phase plane of $([S],[P])$ (Fig.\ref{fig:TURofglycolysis}B). 
A trajectory with the mean oscillatory period of $\sim$ 400 s ($\approx 6-7$ min), fluctuating between $\sim$ 0.1 mM and $\sim$ 1.5 mM, emerges at the condition ($L=4\times 10^9$, $\nu=0.005$ mM/s, $k_s=0.05$ s$^{-1}$, $c=0.01$) (yellow stars in Figs.~\ref{fig:TURofglycolysis}A, B, C). 
The period and amplitude of the oscillation comport well with those observed in yeast and yeast extract \cite{HESS1969,Hess79JChemEd}, in which PFK1 enzymes are oligomerized to an octameric form.

Next, the FP equation for the glycolytic oscillations is obtained from the CME corresponding to Eq.\ref{eqn:glycolysis} with the following force vector and diffusion tensor: 
\begin{align}
\vec{F}&=\begin{bmatrix} 
	\nu-k\alpha q \\
	k\alpha q -k_s[P]    \end{bmatrix}
	+\frac{1}{2\Omega}\begin{bmatrix}
	-k\alpha q-k\alpha\partial_{[S]}q
	+k\alpha\partial_{[P]} q &\\
	-k_s+k\alpha q+k\alpha\partial_{[S]}q-k\alpha\partial_{[P]}q
	\end{bmatrix},\nonumber\\
{\bf D}&=\frac{1}{2\Omega}
\begin{bmatrix}
	\nu+k\alpha q& -k\alpha q\\
	-k\alpha q& k_s [P]+ k\alpha q\end{bmatrix}, 
\end{align}
where the concentration dependences of $\alpha=\alpha([S])$ (Eq.\ref{eqn:alpha_gamma}) and $q=q([S],[P])$ (Eq.\ref{eqn:q}) are omitted for the simplicity of the expression.  
The FP equation with these $\vec{F}$ and ${\bf D}$ is used to 
calculate $J^\text{ss}([S],[P])$, $P^\text{ss}([S],[P])$, and $\dot{S}_\text{tot}$ based on Eq.\ref{eqn:S}. 

Shown in Fig.\ref{fig:TURofglycolysis}C, D are the 2D diagrams of $\dot{S}_\text{tot}$, $\langle T\rangle$, $\langle \delta T^2\rangle$, $J_\text{cycle}$, $A$, and finally $\mathcal{Q}$ as a function of $\nu$ and $k_s$, which are the two experimentally controllable parameters. 
Overall, the correlations between these quantities are not so strong in comparison with those calculated for the Brusselator (see Fig.~S3). 
It is fair to say that the correlation or a trend seen in Brusselator cannot be generalized to other biochemical oscillators. 
At the parameter values, $\nu=0.005$ mM/s$^{-1}$ and $k_s=0.05$ s$^{-1}$, yielding $T=\langle T\rangle \pm \langle\delta T^2\rangle^{1/2}\approx 400\pm 20$ s,  
the uncertainty product is $\mathcal{Q}\simeq 31$. 
Remarkably, while $\mathcal{Q}\simeq 31$ is observed in the vicinity of the lower phase boundary where the entropy production rate is minimal over the relevant phase space (see Fig.\ref{fig:TURofglycolysis}C and the first panel of Fig.\ref{fig:TURofglycolysis}D), $\langle T\rangle \approx 400$ sec (the second panel of Fig.\ref{fig:TURofglycolysis}D) is obtained only at $(\nu,k_s)\simeq (0.005\text{ mM/s},0.01 \text{ s}^{-1})$.  

\section{Discussions}

\begin{figure}
\includegraphics[width=1\linewidth]{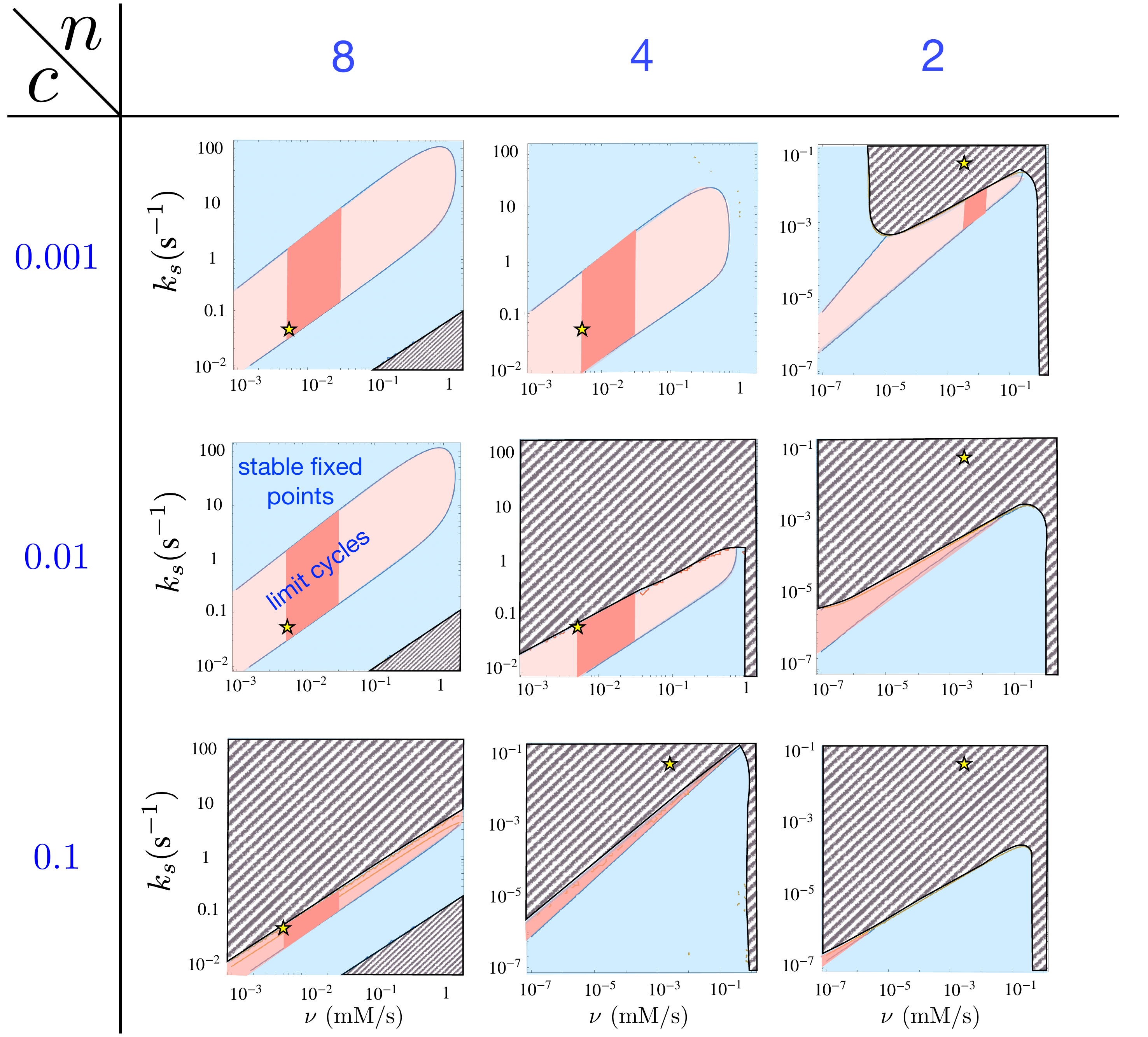}
\caption{The 2D phase diagram of glycolytic oscillations as a function of 
the injection ($\nu$) and degradation rates ($k_s$) for varying $c$ at $n(=2,4,8)$ with $L=4\times10^9$ and $k=500$ s$^{-1}$. 
No catalysis occurs in the hashed region. 
The working condition of glycolytic oscillations ($\nu=0.005$ mM/s, $k_s=0.05$ s$^{-1}$) that produces the oscillations with the mean period $\langle T\rangle =400$ s at $c=0.01$ and $n=8$ (Fig.\ref{fig:TURofglycolysis}) is marked with the yellow star in each panel.   
}
\label{fig:c_n_diagram}
\end{figure}

Glycolytic oscillations are the temporal order that emerges under 
certain special conditions in which parameters defining a set of coupled nonlinear equations yield unstable fixed points. 
Notably, for octameric form of PFK oligomers to demonstrate oscillatory dynamics, the condition of $c=K_R/K_T=0.01$, which renders the substrate binding to the protomer in $R$ state more preferable than to $T$ state by a hundred fold, is essential. If $c$ is increased to 0.1, the phase space corresponding to limit cycles significantly narrows down (Fig.\ref{fig:c_n_diagram}). 
For tetrameric form of PFK ($n=4$), which pertains to bacteria, 
the phase space region for limit cycles is much narrower even when $c=0.01$ and $L$ is set to the value of octamer ($L=4\times 10^9$) (Fig.\ref{fig:c_n_diagram}). 
Unless $k_s$ is tuned to a narrow interval of $k_s\simeq 0.01 - 0.1$ s$^{-1}$, no oscillation is expected. 
The $(\nu,k_s)$ phase diagrams of glycolysis with varying $c$ and $n$ (Fig.\ref{fig:c_n_diagram}) rationalize why glycolytic oscillations were only reported in eukaryotic PFK, where PFK exists in the octameric form.

The TUR, which specifies the physical lower bound to the uncertainty product, is used to assess how the period of temporal order emerging from the underlying dynamical process is balanced with the dissipation under the constraint of cost-precision trade-off. 
In the Brusselator, whose dynamical behavior is defined only with two parameters ($a$ and $b$), there is a specific case that both precision of oscillatory period and dissipation are simultaneously minimized to yield a reasonably small uncertainty product $\mathcal{Q}\simeq 17$ over the phase space. 
In comparison, glycolytic oscillations are more complicated with many more parameters ($L$, $c$, $n$, $K_R^S$, $K_R^P$, $k$, $\nu$, $k_s$). 
To simplify the problem, we have reduced the unknowns by assuming that some of the parameters have identical values with those pertaining to the protomer. 
The values of uncertainty product $\mathcal{Q}$ for the glycolytic oscillations at their working condition producing the period of $\sim$ (5 -- 10) min is  
$\mathcal{Q}\simeq 31$. Remarkably, given the substrate injection rate $\nu=0.005-0.05$ s$^{-1}$, $\mathcal{Q}\simeq 31$ is effectively the minimal value over the phase space involving the limit cycle (Fig.\ref{fig:TURofglycolysis}C). 
$\mathcal{Q}\simeq 31$ is greater than those determined for the molecular motors $\mathcal{Q}\approx 7 - 15$ \cite{Hwang2018JPCL,Mugnai2020RMP}, and biological copy process by exonuclease-deficient T7 DNA polymerase $\mathcal{Q}\approx 10$ \cite{Song2020JPCL}, but smaller than $\mathcal{Q}\approx 45-50$ for the translation process by {\it E. coli} ribosome \cite{Pineros2020PRE,Song2020JPCL}. 
In comparison with the uncertainty product determined for other biochemical cycles ($\mathcal{Q} \approx 10^3$) \cite{marsland2019JRSI}, which severely underperform the TUR's lower bound of $2$, the value of the uncertainty product $\mathcal{Q}\simeq 31$ for the glycolytic oscillation arising from octameric PFK is significantly smaller, minimizing the entropy production rate over the relevant phase, which indicates the cost-effectiveness of the molecular mechanism generating the oscillatory dynamics. 

Lastly, it is of particular note that the $\sim$ (5 -- 10) min oscillation period is observed in cellular or physiological scales as well, such as the blood glucose level, intracellular Ca$^{2+}$ concentrations, and membrane action potentials, which are controlled by the pulsatile secretion of insulin with a period of $\sim$ (5 -- 10) min \cite{mckenna2016PLOS,Lang1979NEJM,WESTERMARK2003BJ}, suggestive of a connection between the dynamics at the molecular and macroscopic scales \cite{bertram2004BJ}, and their synchronization \cite{lee2017PLOS}.  
Remarkably, the loss of pulsatile insulin release resulting from elevated glucose level \cite{mckenna2016PLOS,lee2017PLOS} is also consistent with our study that the working condition of glycolytic oscillation is situated near the borderline of the phase boundary.  
The phase diagram depicted in Fig.\ref{fig:TURofglycolysis}A predicts that a moderate elevation of glucose injection rate beyond $\nu\approx 0.01$ mM/s for a fixed $k_s=0.05$ s$^{-1}$ would abolish the oscillations.  

\acknowledgments{We thank Prof. Junghyo Jo for illuminating discussions on glucose level oscillations.  
This study is supported by KIAS Individual Grants  CG076501 (P.K.) and CG035003 (C.H.). We thank the Center for Advanced Computation in KIAS for providing computing resources.}

\bibliographystyle{apsrev4-1}
\bibliography{reference,mybib1}

\section{Supporting Information}
\setcounter{table}{0}
\renewcommand{\thetable}{S\arabic{table}}%
\setcounter{figure}{0}
\renewcommand{\thefigure}{S\arabic{figure}}%
\setcounter{equation}{0}
\renewcommand{\theequation}{S\arabic{equation}}%

{\bf Non-dimensionalization of rate equations. }
The stochastic version of the Brusselator is written as  
\begin{align}
\frac{dX}{dt}&=k_1A -k_4 X+k_3\frac {X(X-1)Y}{\Omega^2},\nonumber\\
\frac{dY}{dt}&=k_2 B-k_3\frac {X(X-1)Y}{\Omega^2}
\end{align}
where $\Omega$ is the volume of the system. 
Using the following transformations of the variables and parameters,  
\begin{align}
\left(\frac{k_3}{k_4}\right)^{1/2}\frac{X}{\Omega}&\longrightarrow x,\nonumber\\
\left(\frac{k_3}{k_4}\right)^{1/2}\frac{Y}{\Omega}&\longrightarrow y,\nonumber\\ k_4t&\longrightarrow t,\nonumber\\ 
\frac{k_1}{k_4}\left(\frac{k_3}{k_4}\right)^{1/2}\frac{A}{\Omega}&\longrightarrow a,\nonumber\\ 
\frac{k_2}{k_4}\left(\frac{k_3}{k_4}\right)^{1/2}\frac{B}{\Omega}&\longrightarrow b,
\end{align} 
one can write down a non-dimensionalized version of the rate equations at the limit of $\Omega \rightarrow  \infty$ as in the main text (Eq.11). 
\\

{\bf Diffusion approximation: Fokker Planck equation from Chemical Master Equations.}
For a chemical species (X) involved in a reaction:
\begin{equation}
sX + \cdots {\xrightarrow k} s'X + \cdots
\end{equation}
the time evolution equation of $X$ at the deterministic limit can be written as 
\begin{equation}
\frac{d[X]}{dt} = S k[X]^s
\end{equation}
where $[X]$ is the concentration of $X$, $S=s'-s$ is the stoichiometric coefficient. 
The chemical state of the system at any time is fully determined by the state vector ${\bf X} = (X_1,X_2,\dots, X_N)$ where $X_i$ is the number of chemical species $X_i$ in a compartment of a finite volume $\Omega$.
Then, probability distribution for the system to be in state $\textbf{X}$ at time $t$ is
\begin{align}
&P(\textbf{X},t+dt) = P(\textbf{X},t) \nonumber\\
&+dt\sum_{r=1}^R\left[f_r(\textbf{X}- \textbf{S}_r)P(\textbf{X}-\textbf{S}_r,t)-f_r(\textbf{X})P(\textbf{X},t)\right].
\label{eqn:CME1}
\end{align}
where $f_r(\textbf X)=k_r\Omega\prod_{i=1}^N \frac{X_i!}{(X_i-s_{ir})!\Omega^{s_{ir}}}$ is the probability for the reaction $r$ to occur. 
At the limit of $dt\rightarrow 0$, Eq.\ref{eqn:CME1}
is cast into the Chemical Master Equation (CME)
\begin{equation}
{\partial_t}P(\textbf{X},t)=\sum_{r=1}^R \left\{f_r(\textbf{X}- \textbf{S}_r)P(\textbf{X}-\textbf{S}_r,t)-f_r(\textbf{X})P(\textbf{X},t)\right\}.
\end{equation}
Since analytic solutions of CME is known only for limited cases, 
CME is typically approximated to Fokker Planck equation through a Taylor expansion of the relevant terms to the second order,
\begin{align}
f_r(\textbf{X} - \textbf{S}_r)&P(\textbf{X}-\textbf{S}_r,t)-f_r(\textbf{X})P(\textbf{X},t) \nonumber\\
&\approx -\sum_{i=1}^N S_{ir}\partial_{X_i}[f_r(\textbf{X})P(\textbf{X},t)] \nonumber\\
&+\sum_{i,j=1}^N S_{ir}S_{jr}\partial_{X_i}\partial_{X_j}[f_r(\textbf{X})P(\textbf{X},t)], 
\end{align}
leading to 
\begin{align}
{\partial_t}P(\textbf{X},t)&=-\sum_{i=1}^N \partial_{X_i}[A_i(\textbf{X})P(\textbf{X},t)]\nonumber\\
&+\frac 1 2 \sum_{i,j=1}^N \partial_{X_i}\partial_{X_j}[B_{ij}(\textbf{X})P(\textbf{X},t)]
\end{align}
where the drift vector $A$ and diffusion matrix $B$ are given by
\begin{align}
A_i(\textbf{X})&=\sum_{r=1}^R S_{ir}f_r(\textbf{X}),\nonumber\\
B_{ij}(\textbf{X})&=\sum_{r=1}^R S_{ir}S_{jr}f_r(\textbf{X}).
\end{align}
In order to handle the time evolution of different chemical species inside the compartment of volume $\Omega$ in terms of their concentrations, we define 
$x_i\equiv X_i/\Omega\equiv [X_i]$.
With this definition, the probability density of a set of concentrations can be converted to that of molecular counts as $P(\textbf{x},t)\equiv \Omega^N P(\Omega \textbf{x}, t)$. 
Finally, the Fokker-Planck equation for the stochastic time evolution of concentrations of chemical species is obtained as 
\begin{align}
{\partial_t}P(\textbf{x},t)&=-\sum_{i=1}^N \partial_{x_i}[A_i(\textbf{x})P(\textbf{x},t)]\nonumber\\
&+\frac {1}{2\Omega} \sum_{i,j=1}^N \partial_{x_i}\partial_{x_j}[B_{ij}(\textbf{x})P(\textbf{x},t)].
\end{align}

\begin{figure}[t]
\includegraphics[width=0.9\columnwidth]{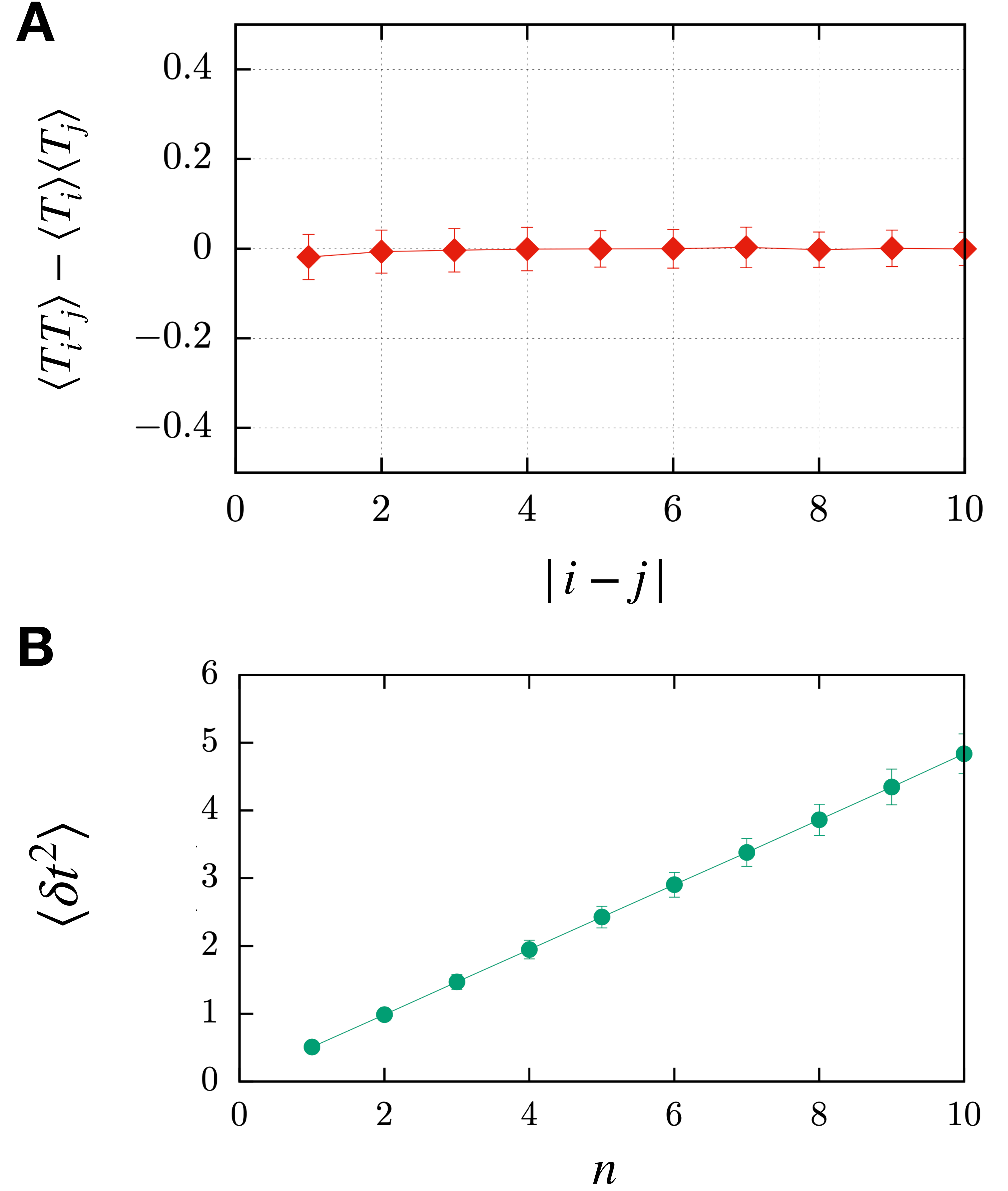}
\caption{{\bf A}. $\langle\delta T_i\delta T_j\rangle=\langle T_i T_j \rangle - \langle T_i\rangle \langle T_j \rangle$ versus $|i-j|$. When $|i-j|\gtrsim 3$, the correlation between two distinct periods is effectively zero ($\langle\delta T_i\delta T_j\rangle\approx 0$). 
{\bf B}. $\langle \delta t^2 \rangle $ versus $n$. 
The plot of $\langle \delta t^2 \rangle $ versus $n$ indicates that the variance of mean first passage time $\langle\delta t^2\rangle$ increases linearly with $n$ (the number of oscillations), satisfying $\langle\delta t^2\rangle=Dn$, where $D=\langle\delta T^2\rangle$. 
}
\label{fig:Tcorrelation}
\end{figure}

\begin{figure}[t]
\includegraphics[width=0.9\linewidth]{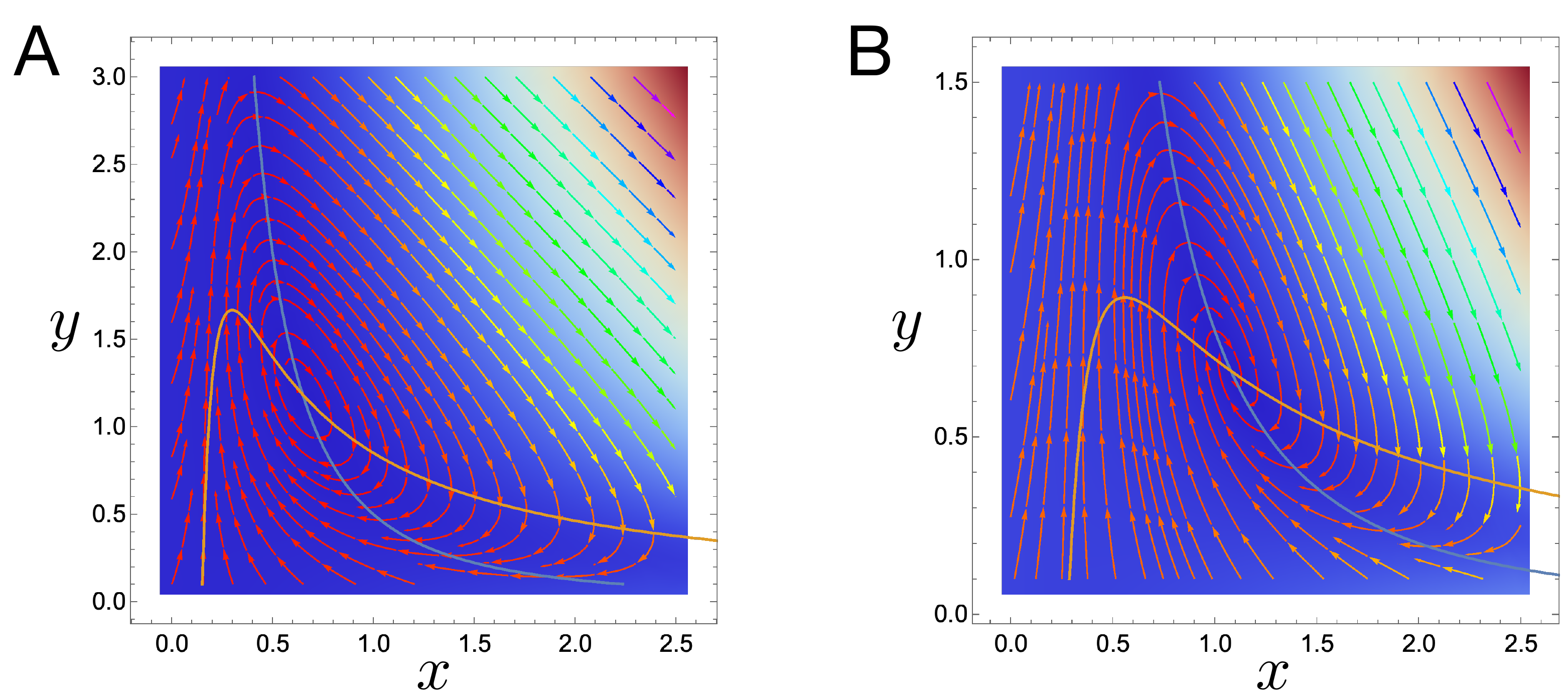}
\caption{Vector field $(dx/dt, dy/dt)$ of Brusselator around the fixed point depicted on the phase plane. 
(A) Unstable fixed point with $a=0.15$ and $b=0.5$.  
(B) Stable fixed point with $a=0.28$ and $b=0.8$. 
$y=1/x-a/x^2$ and $y=b/x^2$ are drawn in pink and pale blue, respectively. 
\label{fig:phaseplane}
}
\end{figure}

\begin{figure}[t]
\includegraphics[width=0.9\linewidth]{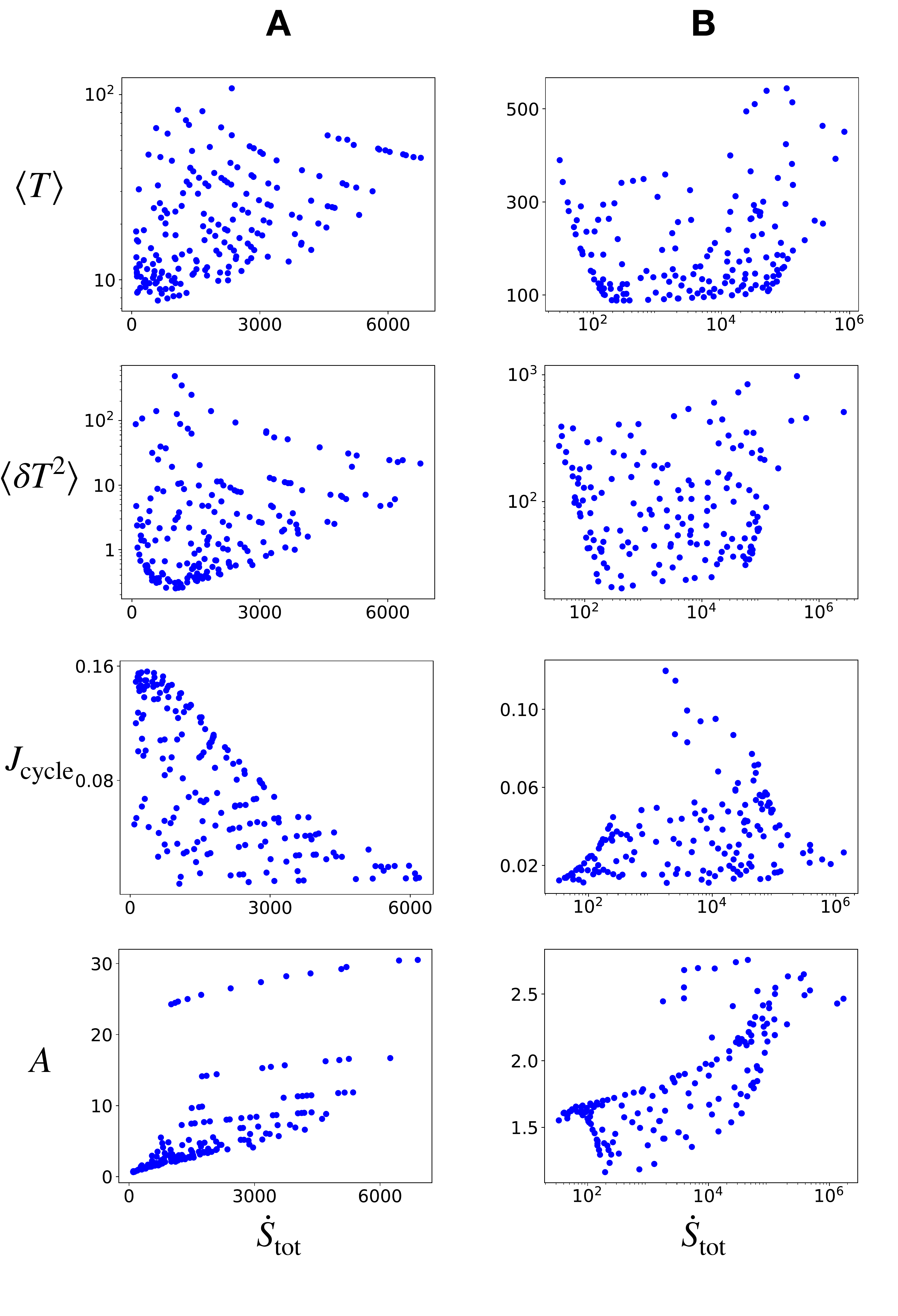}
\caption{Correlations between $\dot{S}_\text{tot}$ and $\langle T\rangle$, $\langle(\delta T)^2\rangle$, $J_\text{cycle}$, $A$ for (A) brusselator and (B) glycolytic oscillator models. The data points were generated by randomly selecting the parameter values pertaining to the phase region of limit cycle. 
\label{fig:SJA}
}
\end{figure}

\begin{figure}
\includegraphics[width=0.9\linewidth]{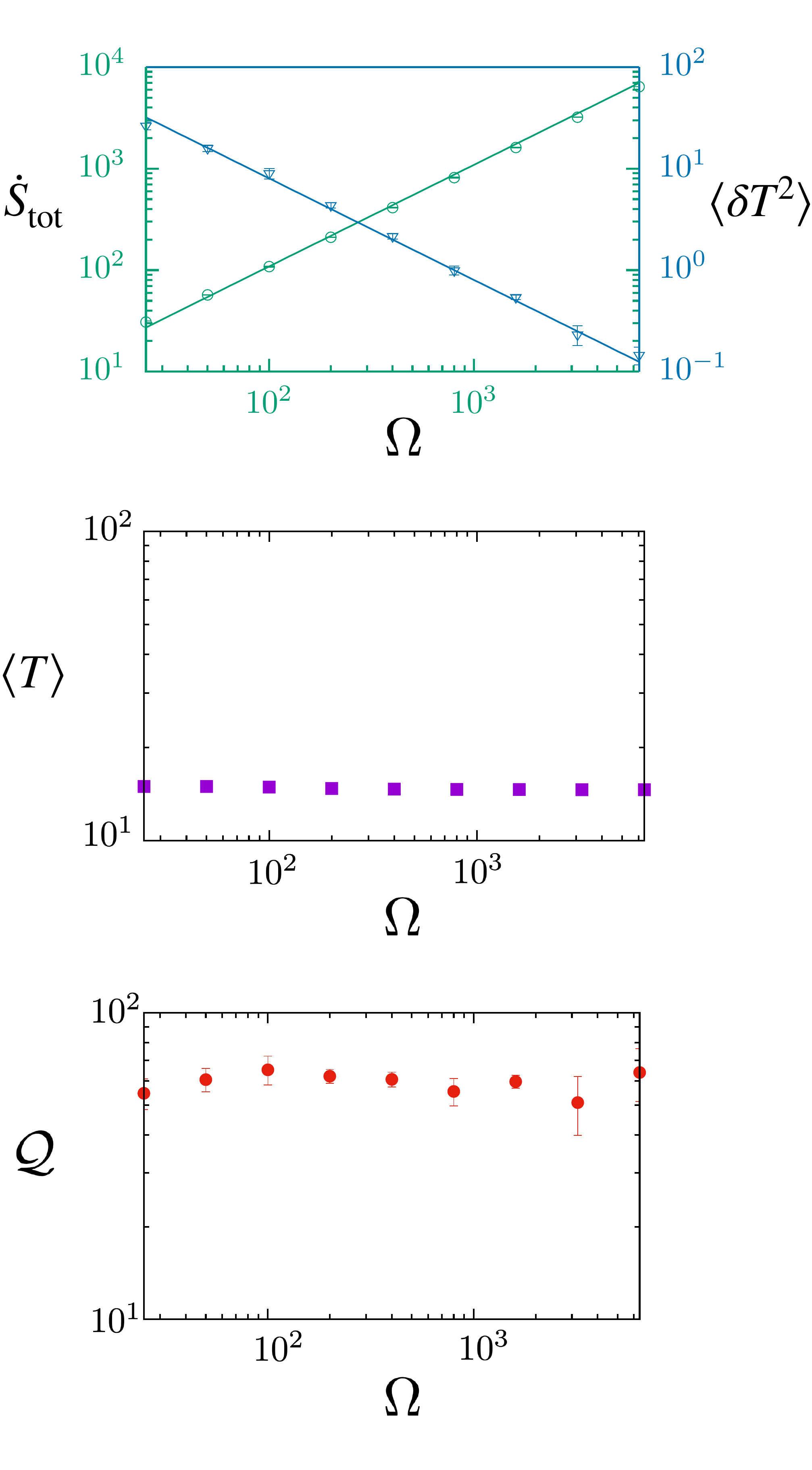}
\caption{Effect of system size on the entropy production, fluctuations, oscillatory period, and $\mathcal{Q}$. 
The data are fitted to $\dot{S}_{\text{tot}}\sim \Omega$ , $\langle\delta T^2\rangle\sim \Omega^{-1}$. 
$\langle T\rangle$ and $\mathcal{Q}$ are effectively constant, independent of $\Omega$.
\label{fig:Omega}
}
\end{figure}

\end{document}